\documentclass[11pt,twoside]{article}

\usepackage{asp2006}
\usepackage{epsf}
\usepackage{psfig}
\usepackage{lscape}

\begin{document}
\title{Revealing an Energetic Galaxy-Wide Outflow in a $z\approx$~2 Ultraluminous Infrared Galaxy}   
\author{D.~M.~Alexander$^1$, A.~M.~Swinbank$^2$, I.~Smail$^2$, R.~McDermid$^3$, and N.~P.~H.~Nesvadba$^4$}   
\affil{$^1$Department of Physics, Durham University, Durham, DH1 3LE, UK\\
$^2$Institute for Computational Cosmology, Durham University, Durham, DH1 3LE, UK\\
$^3$Gemini Observatory, 670 N.~\ A`ohoku Place, Hilo, HI 96720, USA\\
$^4$Institut d'Astrophysique Spatiale, Universit\'e Paris Sud 11, Orsay, France
}    

\begin{abstract} 
Leading models of galaxy formation require large-scale energetic
outflows to regulate the growth of distant galaxies and their central
black holes. However, current observational support for this
hypothesis at high redshift is mostly limited to rare $z>2$ radio
galaxies. Here we present Gemini-North NIFS Intregral Field Unit (IFU)
observations of the [O~{\sc iii}]$\lambda$5007 emission from a
$z\approx$~2 ultraluminous infrared galaxy (ULIRG; $L_{\rm
IR}>10^{12}$~$L_{\odot}$) with an optically identified Active Galactic
Nucleus (AGN). The spatial extent ($\approx$~4--8~kpc) of the high
velocity and broad [O~{\sc iii}] emission are consistent with that
found in $z>2$ radio galaxies, indicating the presence of a
large-scale energetic outflow in a galaxy population potentially
orders of magnitude more common than distant radio galaxies. The low
radio luminosity of this system indicates that radio-bright jets are
unlikely to be responsible for driving the outflow. However, the
estimated energy input required to produce the large-scale outflow
signatures (of order $\approx10^{59}$~ergs over $\approx$~30~Myrs)
could be delivered by a wind radiatively driven by the AGN and/or
supernovae winds from intense star formation. The energy injection
required to drive the outflow is comparable to the estimated binding
energy of the galaxy spheroid, suggesting that it can have a
significant impact on the evolution of the galaxy. We argue that the
outflow observed in this system is likely to be comparatively typical
of the high-redshift ULIRG population and discuss the implications of
these observations for galaxy formation models.
\end{abstract}


\section{Introduction}

Modern-day cosmology is a rich cocktail of observations and
theory. Hard observational constraints guide the theoretical models,
which then provide more detailed insights into the potential physical
mechanisms that drive the growth of galaxies and their massive central
black holes. One area where theoretical models have been particularly
successful is in highlighting that star formation must have been
truncated in the most massive galaxies at high redshifts
(\hbox{$z\approx$~2--3;} e.g.,\ Benson et~al. 2003; Di Matteo
et~al. 2005; Bower et~al. 2006). The leading candidates to cause this
truncation are galaxy-wide outflows and winds, associated with either
star-formation activity (e.g.,\ supernovae winds) or active galactic
nuclei (AGNs; e.g.,\ winds and jets initiated from the black-hole
accretion disk). If energetic enough, these winds could heat the cool
galactic gas and/or eject it from the gravitational potential of the
host galaxy, effectively shutting down any further significant star
formation.

Integral field units (IFUs) are the ideal tool to identify galaxy-wide
energetic outflows as they provide spectro-imaging over several
arcsecond fields of view (e.g.,\ Swinbank et~al. 2005, 2006). The
typical expected signature of an outflow is a broad emission-line gas
component kinematically distinct from the narrow-emission line gas in
the host galaxy (velocity offsets of several hundreds of
kilometers/second) that is extended over kpc-scales. IFU observations
of the [O~{\sc iii}] (rest-frame 5007~ang) emission in a handful of
massive radio-loud AGNs at $z\approx$~2--3 have indeed revealed the
signatures of galaxy-scale energetic outflows (e.g.,\ Nesvadba
et~al. 2006, 2007, 2008). The outflows from these systems appear to be
driven by radio jets initiated by AGN activity. While these data
provide evidence that energetic outflows can be identified in distant
galaxies, radio-loud AGNs are rare beasts and it is far from clear how
typical these outflows are in the typical distant massive galaxy
population. To throw light on whether outflows are ubiquitous in the
distant Universe, we have used the Gemini NIFS IFU to search for
galaxy-wide energetic outflows in more typical radio-quiet
systems. Here we present our initial results from this program on the
$z\approx$~2 radio-quiet AGN SMM~J1237+6203, which are published in
Alexander et~al. (2009).

SMM~J1237+6203 is an optically bright $z=$~2.07 quasar that is also
bright at submillimetre (850~$\mu$m), radio (1.4~GHz), and X-ray
(0.5--8 keV) wavelengths (Alexander et~al. 2003, 2005; Barger
et~al. 2003; Chapman et~al. 2005). The estimated infrared luminosity
of SMM~J1237+6203 is of order $6\times10^{12}$~$L_{\odot}$, indicating
that it is a distant ultra-luminous infrared galaxy (ULIRG). The AGN
in SMM~J1237+6203 is luminous (X-ray luminosity
$\approx10^{44}$~erg~s$^{-1}$) and undoubtably contributes to a
considerable fraction of the infrared luminosity (Alexander
et~al. 2005). However, given the bright submillimetre and radio
emission, it seems likely that SMM~J1237+6203 also hosts an
ultra-luminous infrared starburst in addition to the luminous AGN
activity. Previously published near-infrared spectroscopy shows that
SMM~J1237+6203 has bright, and possibly broad and extended, [O~{\sc
iii}] emission (Takata et~al. 2006). These are the expected signatures
of a galaxy wide outflow, although high-quality IFU observations are
required to provide detailed constraints.

\setcounter{figure}{0}
\begin{figure}
\plotfiddle{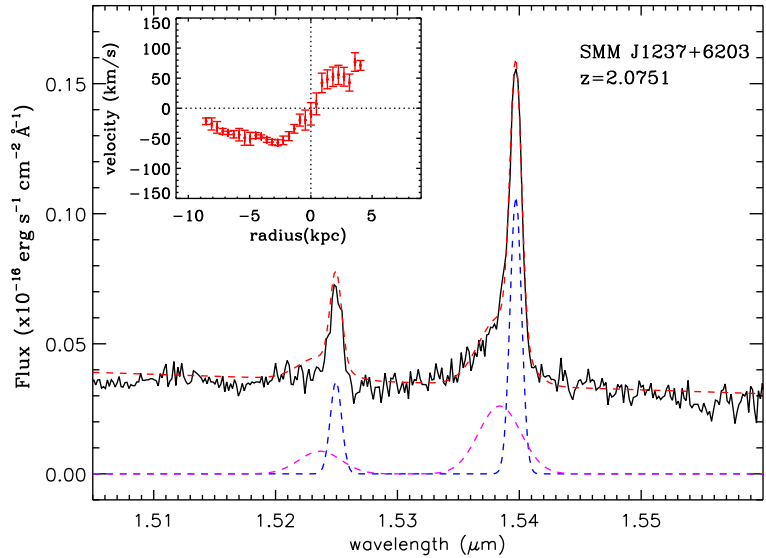}{2.75in}{90.}{45.}{45.}{170}{-30}
\caption{Collapsed one-dimensional NIFS spectrum showing the emission
line profiles fitted with both a broad and narrow emission-line
component. Inset plot shows the narrow [O~{\sc iii}] velocity
field. Taken from Alexander et~al. (2009).}
\end{figure}

\section{Gemini NIFS IFU Observations}

We obtained the NIFS IFU observations of SMM~J1237+6203 on 2008 May 22
and 2008 May 30, with a total on-source integration time of 7.8~ks
(600s individual exposures obtained in a standard ABBA
configuration). The observations were taken in excellent photometric
conditions with $\approx$~0.3 arcsec seeing. The NIFS IFU uses an
image slicer to take a $3\times3$ arcsec field with a pixel scale of
0.043 arcsec and divides it into 29 slices of width 0.103 arcsecs. The
collapsed one-dimensional spectrum is shown in Fig 1. The [O~{\sc
iii}] emission line has a blue asymmetric profile, which is fitted
with two underlying Gaussian profiles: the offset between the narrow
emission line (FWHM~$\approx$~210~km~s$^{-1}$) and the broad emission
line (FWHM~$\approx$~820~km~s$^{-1}$) is
$\Delta$V$\approx$~--250~km~s$^{-1}$. The luminosity of the broad and
narrow components are comparable ($\approx10^{43}$~erg~s$^{-1}$). The
luminous [O~{\sc iii}] emission is the result of photoionisation by
the AGN, although it is less certain what is responsible for the
production of the kinematically complex [O~{\sc iii}] component.

To provide spatial information of the [O~{\sc iii}] emission, we
constructed intensity, velocity, and FWHM maps of the [O~{\sc iii}]
emission from the IFU data cube. A $\chi^2$ minimisation procedure was
used to fit each spectrum within the datacube, taking into account the
greater noise at the positions of the sky lines. The spectra were
averaged in increasingly larger spatial bins until a significant
emission-line component was identified. To detect an emission line we
required a $>5$~$\sigma$ improvement over a simple continuum fit, and
to detect an additional broad component we required a $>4$~$\sigma$
improvement over a single narrow emission-line fit. The basic [O~{\sc
iii}] properties derived from the IFU datacube are shown in Fig 2.

\setcounter{figure}{1}
\begin{figure}
\plotfiddle{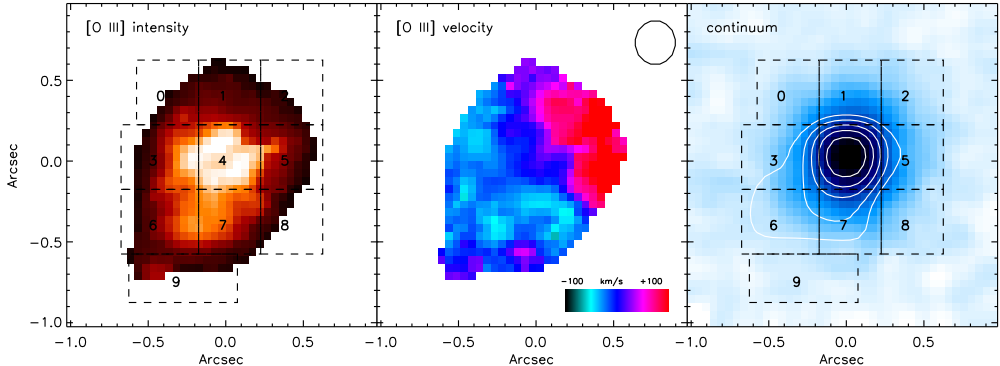}{2.2in}{90.0}{50.}{50.}{200}{0}
\caption{[O~{\sc iii}] intensity (left), narrow [O~{\sc iii}] velocity
map (middle), and line-free continuum image (right); circle in top
right denotes the seeing disk size. The contours represent an
intensity weighted map of the broad [O~{\sc iii}] emission
components. The numbered regions of the broad [O~{\sc iii}] emission
are plotted in Fig.~3. Taken from Alexander et~al. (2009).}
\end{figure}

\section{Evidence for an Energetic Galaxy-Wide Outflow}

Narrow [O~{\sc iii}] emission is detected across $\approx$~14~kpc,
which is blue-shifted and red-shifted (with respect to the systemic
redshift) to the South-East and North-West of the nucleus,
respectively; see Fig.~2. The velocity field of this narrow [O~{\sc
iii}] emission may be dominated by the host-galaxy rotation, although
it is not clear that this is the only plausible explanation; see Fig
1. The spheroid mass estimated from the velocity dispersion of the
narrow [O~{\sc iii}] emission ($\approx$~200~km~s$^{-1}$) is of order
$\approx10^{11}$~$M_{\odot}$, and is consistent with that expected
from the estimated black-hole mass ($M_{\rm
BH}\approx2\times10^8$~$M_{\odot}$; Alexander et~al. 2008), given the
local black-hole--spheroid mass relationship.

Broad [O~{\sc iii}] emission is detected over $\approx$~4--8~kpc at
the nucleus and to the South-East of the nucleus; see Fig.~2. The
radial extent, velocity offset, and FWHM of the broad [O~{\sc iii}]
emission is shown in Fig.~3. The broadest [O~{\sc iii}] components
clearly correspond to those with the largest velocity offset. These
characteristics are similar to those found in some distant radio-loud
AGNs (e.g.,\ Nesvadba et~al. 2006, 2007, 2008) and are consistent with
those expected for an energetic galaxy-wide outflow. Assuming that the
broad [O~{\sc iii}] emission is due to an energy-conserving bubble
expanding into a uniform region, the kinetic energy required to
produce the broad [O~{\sc iii}] features is of order
$\approx$~(0.6--3)~$\times10^{44}$~erg~s$^{-1}$ over 4--8~kpc
(calculated using Eqn.~3 in Nesvadba et~al. 2006). Over a typical
AGN/starburst lifetime of $\approx$~30~Myrs, the total injection of
energy into the outflow would be of order
$\approx$~(0.3--3)~$\times10^{59}$~ergs, which is comparable to the
estimated binding energy of the galaxy spheroid
($\approx10^{59}$~erg~s$^{-1}$). This analysis is based on a simple
model and should only be considered illustrative with uncertainties at
the level of an order of magnitude but, given the limited constraints
available for high-redshift systems, a more complex model is not yet
warranted. However, it does indicate that the large-scale outflow in
SMM~J1237+6203 may be energetic enough to unbind at least a fraction
of the gas from the host galaxy.

\setcounter{figure}{2}
\begin{figure}[!t]
\plotone{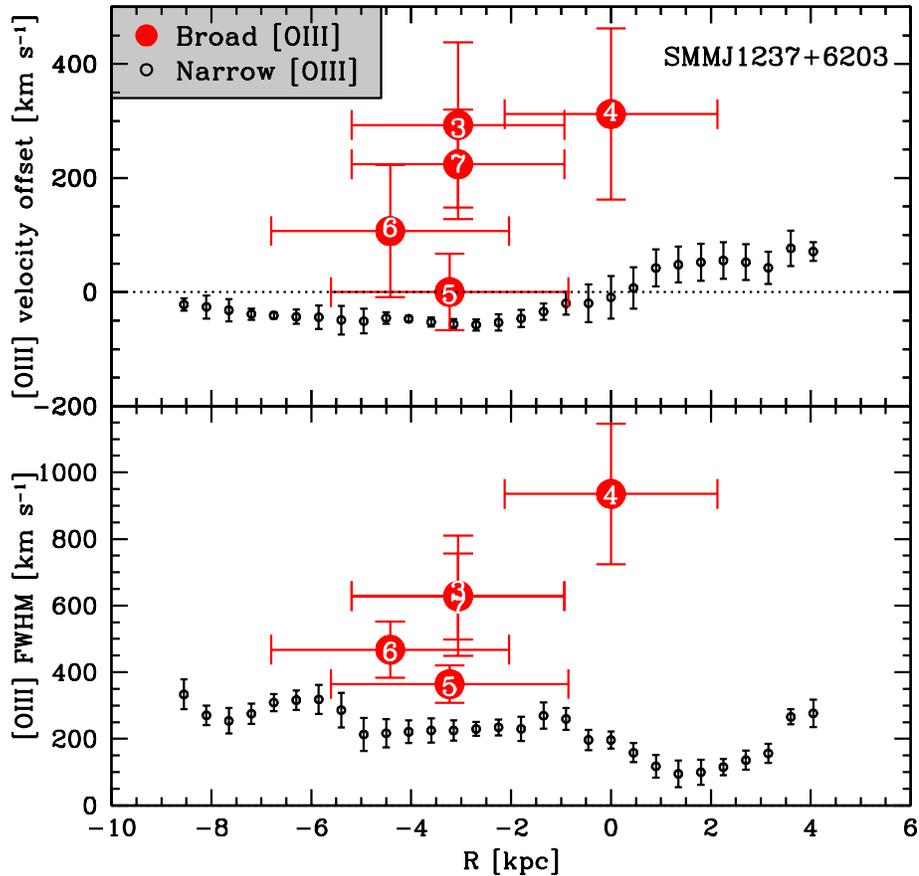}
\caption{Velocity and FWHM of the broad and narrow [O~{\sc iii}]
emission plotted as a function of radius from the nucleus. The
numbered regions correspond to those indicated in Fig.~2. Taken from
Alexander et~al. (2009).}
\end{figure}

What is driving this large-scale energy outflow? Both AGN activity and
star formation are plausible candidates. Assuming that $\approx$~10\%
of the mass accreted onto the black hole is also liberated as an
accretion-disk outflow (as motivated by observations of similar
distant AGNs; e.g.,\ Chartas et~al. 2007), the initial energy input
into the accretion-disk wind would be of order
\hbox{$\approx$~(0.3--3)~$\times10^{45}$~erg~s$^{-1}$}. Therefore, so
long as $\approx$~2--100\% of this energy can be coupled to the
host-galaxy gas, an accretion-disk wind in SMM~J1237+6203 could drive
the outflow over $\approx$~4--8~kpc. Similarly, assuming a typical
star-formation rate for submillimeter-emitting galaxies of
$\approx$~1000~$M_{\odot}$~yr$^{-1}$, the predicted energy injection
from supernovae winds would be of order
$\approx3\times10^{44}$~erg~s$^{-1}$. If $\approx$~20--100\% of this
energy can be coupled to the host-galaxy gas, then star-formation
activity could also drive the large-scale outflow. We cannot
distinguish between these two different scenarios on the basis of the
current data. However, given the comparatively modest radio luminosity
from SMM~J1237+6203, we can rule out that the outflow if driven by
radio jets, in contrast to that found in distant radio-loud AGNs; the
gas coupling efficiency in SMM~J1237+6203 would need to be
unphysically high ($\approx$~10,000\%), assuming typical radio-jet
models. See \S4 of Alexander et~al. (2009) for more details.

SMM~J1237+6203 is the first high-redshift ULIRG with spatially
extended broad [O~{\sc iii}] emission to be mapped with IFU
observations. Previous IFU studies of high-redshift ULIRGs have
typically focused on the Ly~$\alpha$ or H~$\alpha$ emission line, and
the majority of the objects observed have not shown evidence for AGN
activity at optical wavelengths (e.g.,\ Swinbank et~al. 2005,
2006). However, three pieces of indirect evidence suggest that
SMM~J1237+6203 could be relatively typical of the high-redshift ULIRG
population: (1) broad [O~{\sc iii}] emission-line components have been
identified with rest-frame optical spectroscopy in several
high-redshift ULIRGs to date (e.g.,\ Takata et~al. 2006), (2)
$\approx$~50\% of nearby ULIRGs hosting optical AGN activity have
[O~{\sc iii}] components with FWHM~$>$~800~km~s$^{-1}$, comparable to
that found for SMM~J1237+6203 (e.g.,\ Veilleux et~al. 1999; Zheng
et~al. 2002), and (3) high-quality IFU data have been published for a
number of nearby ULIRGs showing that they host broad and extended
[O~{\sc iii}], providing evidence for large-scale energetic outflows
in at least some ULIRGs in the local Universe (e.g.,\ Wilman
et~al. 1999; L{\'{\i}}pari et~al. 2009).

It therefore seems likely that the outflow processes seen in
SMM~J1237+6203 are comparatively common in the distant Universe. Since
ULIRGs are orders of magnitude more common than radio-loud AGNs at
$z\approx$~2, their global contribution to the injection of energy
into the host galaxy could therefore be very significant and even
dominant. The NIFS IFU observations for the other targets in our
distant AGN--ULIRG sample should provide at least the first steps
toward addressing this issue.


\acknowledgements 

We would like to thank the following organisations for support: Royal
Society (DMA), the Philip Leverhulme Prize Fellowship (DMA), the Royal
Astronomical Society (AMS), and the Science and Technology Facilities
Council (STFC).


\end{document}